\documentclass[11pt,a4paper]{article}
\usepackage{authblk}
\usepackage[utf8]{inputenc}
\usepackage[round]{natbib}
\usepackage[pdftex]{graphicx}
\usepackage{amsmath,amsthm,amsfonts}
\usepackage{lscape}   
\usepackage{rotating}
\usepackage{rotfloat}
\usepackage{multirow}
\usepackage{booktabs}
\usepackage{ctable}
\usepackage{threeparttable}
\usepackage{booktabs}
\usepackage{framed}
\usepackage{color}
\usepackage[colorlinks=true,linkcolor=blue,allcolors=blue]{hyperref}

\usepackage{soul}

\title{\Large \bf On Bayesian wavelet shrinkage estimation of nonparametric regression models with stationary errors \thanks{The second author acknowledges the financial support of the S\~ao Paulo State Research Foundation (FAPESP), grant 2023/02538-0 and the support of the  the Centre for Applied Research on Econometrics, Finance and Statistics (CAREFS).}}

\author{Alex Rodrigo dos S. Sousa}
\author{Mauricio Zevallos}

\affil{Universidade Estadual de Campinas (UNICAMP)\\ Departament of Statistics, Brazil \thanks{ Sousa (asousa@unicamp.br) and Zevallos (amadeus@unicamp.br)}}

\date{\today}

\begin{document}

\maketitle

\begin{abstract}
This work proposes a Bayesian rule based on the mixture of a point mass function at zero and the logistic distribution to perform wavelet shrinkage in nonparametric regression models with stationary errors (with short or long-memory behavior). The proposal is assessed through Monte Carlo experiments and illustrated with real data. Simulation studies indicate that the precision of the estimates decreases as the amount of correlation increases. However, given a sample size and error correlated noise, the performance of the rule is almost the same while the signal-to-noise ratio decreases, compared to the performance of the rule under independent and identically distributed errors. Further, we find that the performance of the proposal is better than the standard soft thresholding rule with universal policy in most of the considered underlying functions, sample sizes and signal-to-noise ratios scenarios.\\


\end{abstract}

\newpage
\section{Introduction}

In nonparametric regression problems, the usual approach is to represent the unknown function in terms of a linear combination of some functional basis. In this way, the statistical infinite dimensional problem of estimating the unknown function becomes a finite dimensional problem of estimating the coefficients of the representation. There are several functional basis widely applied in this regard such as polynomials, splines and their expansions, Fourier basis, and wavelets, the focus of this work. The representation of a function in wavelet basis brings several advantages in terms of its coefficients. In fact, they are well localized in both time and frequency domains, i.e, significant wavelet coefficients are found in important features of the function such as peaks, discontinuities and oscillations. Furthermore, the coefficient vector is typically sparse, i.e, the wavelet coefficients located in smooth regions of the function are equal to or very close to zero, which allows storing the main characteristics of the represented function in terms of a few number of nonzero coefficients, a welcome property from the computational and statistical estimation points of view. See \cite{vidakovic-1999} for more details about wavelet applications in statistical models, and \cite{daubechies-1992} and \cite{mallat-1998} for theoretical developments of wavelets.

Due to the sparsity property of the wavelet coefficients, their estimation is usually done by wavelet shrinkage estimators. After the application of a discrete wavelet transform (DWT) in the original data, the empirical wavelet coefficients are reduced by a wavelet shrinkage in order to decrease their magnitudes, mainly if they are sufficiently close to zero. The shrunk versions of the empirical coefficients are then the estimates of the wavelet coefficients of the functional representation. The function estimate is obtained by the application of the inverse discrete wavelet transform (IDWT). Several wavelet shrinkage rules have been proposed since the seminal works of \cite{donoho-johnstone-1994} and \cite{donoho-johnstone-1995}, who proposed shrinkage rules based on thresholding, the so called soft and hard thresholding rules. See for example \cite{abra-1996} and \cite{nason-1996} for proposed policies to perform wavelet thresholding based on the false discovery rate and cross-validation method respectively and also \cite{vidakovic-1999} and \cite{nason-2008} for a general overview of wavelet shrinkage rules. Bayesian wavelet shrinkage rules have also been considered by researches, since they allow the incorporation of several features of the coefficients, such as the sparsity property and their support, if they are bounded. For instance, see \cite{Chipman-et-al-1997}, \cite{vidakovic-ruggeri-2001}, \cite{angelini-vidakovic-2004}, \cite{remenyi-vidakovic-2015}, \cite{sousa-et-al-2020}, \cite{sousa-2020} and \cite{vimalajeewa-et-al-2023} for a Bayesian shrinkage rules review.

However, most of the proposed wavelet shrinkage rules were obtained and evaluated under the assumption of independent and identically distributed (IID) and zero mean normal errors. Since correlated noises arise in applications, the study of the performance of shrinkage rules under this scenario is important. For example \cite{wang-1996} proposed a fractional Gaussian noise model to deal with nonparametric regression with long-range dependency and established asymptotic properties for the minimax risk. \cite{johnstone-silerman-1997} proposed a level-dependent threshold choice to be applied in the Donoho-Johnstone soft thresholding rule when the data have stationary correlated noise. \cite{porto-et-al-2016} also studied wavelet shrinkage under random design and correlated noise also by adapting the soft thresholding rule  and \cite{morettin-porto-2022} surveyed the estimation of nonparametric regression models using
wavelets. Even though the studies above considered the thresholding rule in nonparametric regression models under correlated noise, little attention has been given to Bayesian shrinkage rules in this context. The present paper addresses this matter.

We propose a level-dependent Bayesian shrinkage rule based on a prior mixture of a point mass function at zero and the symmetric around zero logistic distribution to perform wavelet shrinkage in data with stationary correlated noise, specifically, under an autoregressive process of order one (AR(1)) and autoregressive fractionally integrated moving-average (ARFIMA) noises. This proposal is an extension of the Bayesian shrinkage rule under logistic prior proposed by \cite{sousa-2020} for the case of IID Gaussian noise. The method takes advantage of the logistic distribution which is very suitable as a prior to the wavelet coefficients, since it is unimodal and its hyperparameter controls the degree of shrinkage to be imposed on the empirical coefficients. The proposal is assessed by Monte Carlo experiments involving the Donoho-Johnstone test functions perturbed with correlated noises. In addition, the proposal is compared to the widely known standard soft thresholding rule of \cite{donoho-johnstone-1994}.

The remainder of the paper is organized as follows. In Section \ref{sec:model} we present the model. Section \ref{sec:estimation} describes the proposal estimation procedure. This proposal is assessed by Monte Carlo experiments in Section \ref{sec:simulations}. An illustration is presented in Section \ref{sec:illustration}, and conclusions and final remarks are given in Section \ref{sec:conclusions}. Additional results can be found in the Supplementary Material.

\section{Statistical Model}\label{sec:model}

We consider $n=2^J$ observations $(x_1,y_1),\cdots,(x_n,y_n)$ from the unidimensional nonparametric regression model
\begin{equation}\label{timemodel}
    y_i = f(x_i) + e_i,
\end{equation}
where $x_i$'s are scalars, $f$ is an unknown squared integrable function, i.e, $f \in \mathbb{L}^2(\mathbb{R}) = \{f:\int f^2 < \infty \}$, and $e_i$ denotes random errors that are assumed to be generated by a stationary processes with zero mean and finite variance $\sigma^2_e$.

In this paper we consider three classes of stationary processes for $\{e_i\}$. First, independent and identically distributed (IID) processes with Gaussian distribution. Second, autoregressive processes of order one, AR(1), defined by
\begin{equation}\label{eq:model-ar1}
e_i = \phi e_{i-1} + \eta_i, 
\end{equation}
where the parameter $\phi \in (-1,1)$, and $\{\eta_i\}$ is an IID sequence with Gaussian distribution, $\eta_i \sim N(0, \sigma^2_{\eta})$. Third, autoregressive fractionally integrated moving-average (ARFIMA) processes of order $(0,d,0)$ defined by,
\begin{equation}\label{eq:model-arfima}
(1-B)^d e_i = \eta_i, 
\end{equation}
where $B$ is the backshift operator, $(1-B)^d=\sum_{j=0}^\infty b_j(d)B^j$ with $b_j(d) = \Gamma(j-d)/(\Gamma(j+1)\Gamma(-d))$ for $j=0,1,\ldots$ being $\Gamma(\cdot)$ the gamma function, and $0<d<0.5$. ARFIMA processes were introduced by \cite{granger-joyeux-1980} and \cite{hosking-1981} to reproduce long-memory behavior, i.e., significant autocorrelations at high lags, in contrast to short-memory models such as the autoregressive models that exhibit autocorrelations with exponential decay.

On the other hand, the nonparametric standard procedure is to represent the unknown function $f$ in terms of some suitable functional basis. In this work, we expand the function $f$ in \eqref{timemodel} in a wavelet basis
\begin{equation} \label{expan}
f(x) = \sum_{j,k \in \mathbb{Z}}\theta_{j,k} \psi_{j,k}(x), 
\end{equation}
where $\{\psi_{j,k}(x) = 2^{j/2} \psi(2^j x - k),j,k \in \mathbb{Z} \}$ is an orthonormal wavelet basis for $\mathbb{L}^2(\mathbb{R})$ constructed by dilations $j$ and translations $k$ of a function $\psi$ called a wavelet or mother wavelet and $\theta_{j,k}$ are wavelet coefficients that describe features of $f$ at spatial locations $2^{-j}k$ and scales $2^j$ or resolution levels $j$. Thus, according to the representation \eqref{expan}, the problem of estimating the function $f$ is reduced to the problem of estimating a finite number $n$ of wavelet coefficients $\theta_{j,k}$. For simplicity, the subindices $j,k$ will be dropped in the text without loss of interpretation. 

We can rewrite the model \eqref{timemodel} in vector notation as
\begin{equation}\label{vectormodel}
\boldsymbol{y} = \boldsymbol{f} + \boldsymbol{e},
\end{equation} 
where $\boldsymbol{y} = [y_1,\cdots,y_n]'$, $\boldsymbol{f} = [f(x_1),\cdots,f(x_n)]'$ and $\boldsymbol{e} = [e_1,\cdots,e_n]'$. We apply a discrete wavelet transform (DWT) of the original data to change them to the wavelet domain. Although the DWT is usually performed by fast algorithms such as the pyramidal algorithm, it is possible to represent it by a transformation matrix $\boldsymbol{W}$ with dimension $n \times n$ which is applied on both sides of \eqref{vectormodel}. Since the DWT is linear, we obtain the following model in the wavelet domain
\begin{equation}\label{waveletmodel}
\boldsymbol{z} = \boldsymbol{\theta} + \boldsymbol{\varepsilon},
\end{equation}
where $\boldsymbol{z} = \boldsymbol{W}\boldsymbol{y} = [z_1, \cdots, z_n]'$ is the vector of empirical (observed) wavelet coefficients, $\boldsymbol{\theta} = \boldsymbol{W}\boldsymbol{f} = [\theta_1, \cdots, \theta_n]'$ is the sparse vector of unknown wavelet coefficients of $f$ and $\boldsymbol{\varepsilon} = \boldsymbol{W}\boldsymbol{e} = [\varepsilon_1,\cdots,\varepsilon_n]'$ is the vector of random errors. Thus, we can consider the empirical wavelet coefficients $\boldsymbol{z}$ as noisy versions of the unknown wavelet coefficients  $\boldsymbol{\theta}$. For more details about wavelet transforms in statistical models see \cite{vidakovic-1999}.

\section{Wavelet shrinkage rule and estimation}\label{sec:estimation}

The estimation of the vector of wavelet coefficients $\boldsymbol{\theta}$ is done coefficient by coefficient under a Bayesian framework. In this sense, a prior distribution is assigned to a single wavelet coefficient $\theta$, which incorporates our knowledge about its sparsity and symmetry around zero. Then, we consider the prior distribution $\pi(\cdot;\alpha,\tau)$ based on a mixture of a point mass function at zero $\delta_0(\cdot)$ and a symmetric (around zero) logistic distribution $g(\cdot;\tau)$ as proposed by \cite{sousa-2020}:
\begin{equation}\label{prior}
\pi(\theta;\alpha,\tau) = \alpha \delta_0(\theta) + (1-\alpha)g(\theta;\tau),
\end{equation}   
where
\begin{equation}
g(\theta;\tau) = \frac{\exp\{-\theta/\tau\}}{\tau(1 + \exp\{-\theta/\tau\})^2}\mathbb{I}_{\mathbb{R}}(\theta), \nonumber
\end{equation}
$\alpha \in (0,1)$, $\tau > 0$ and $\mathbb{I}_{\mathbb{R}}(\cdot)$ is the indicator function on the real set $\mathbb{R}$. Thus, the prior distribution has two hyperparameters to be elicited, $\alpha$ and $\tau$. These hyperparameters control the severity of the shrinkage imposed by the Bayesian rule on the empirical coefficients. The empirical coefficients are shrunk more with
higher values of $\alpha$ and smaller values of $\tau$ since the prior distribution becomes more concentrated around zero. \cite{sousa-2020} suggested values of $\tau \leq 10$ and the elicitation of $\alpha$ according to the level-dependent proposal of \cite{angelini-vidakovic-2004},
\begin{equation}\label{eq:alpha}
\alpha = \alpha(j) = 1 - \frac{1}{(j-J_{0}+1)^\gamma},
\end{equation}
where $J_ 0 \leq j \leq J-1$, $J_0$ is the primary resolution level, $J$ is the number of resolution levels, $J=\log_{2}(n)$ and $\gamma > 0$. They also suggested that in the absence of additional information, $\gamma = 2$ can be adopted.

\begin{figure}{h} 
    \centering
    \includegraphics[width=0.7\linewidth]{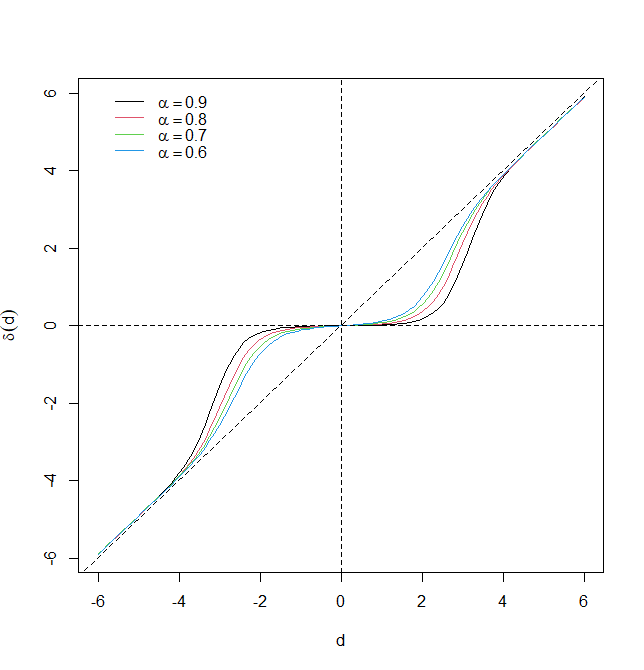}
    \caption{Shrinkage rules \eqref{rule} under the prior \eqref{prior} for $\sigma = 1$, $\tau = 5$ and for $\alpha \in \{0.6, 0.7, 0.8, 0.9\}$.}
    \label{fig:rules}
\end{figure}

To obtain the Bayesian shrinkage rule $\delta(\cdot)$ under the prior \eqref{prior},  we assume the quadratic loss function $L(\delta,\theta) = (\delta - \theta)^2$. Then, Bayes rule gives the expected value of $\theta|z$, i.e, $\delta(z) = \mathbb{E}_{\pi}(\theta|z)$.\footnote{See \cite{robert-2007} for a complete description of Bayesian procedures.} In this paper, in order to account for level dependency that arises for stationary errors, we propose the following version of the shrinkage rule of \cite{sousa-2020}, 
\begin{equation}\label{rule}
\delta_j(z)= \frac{(1-\alpha)\int_\mathbb{R}(\hat{\sigma}_j u + z)g(\hat{\sigma}_j u + z ; \tau)\phi(u)du}{\frac{\alpha}{\hat{\sigma}_j}\phi(\frac{z}{\hat{\sigma}_j})+(1-\alpha)\int_\mathbb{R}g(\hat{\sigma}_j u + z ; \tau)\phi(u)du},
\end{equation}
where $\phi(\cdot)$ is the standard normal density function and $\hat{\sigma}_j$ is the estimate of the standard deviation of the empirical wavelet coefficients at resolution level $j$,
\begin{equation}\label{sigma}
    \hat{\sigma}_j = \mathrm{MAD}\{z_{jk},k=1,\cdots,2^j\}/0.6745,
\end{equation}
where MAD denotes the median absolute deviation. This MAD estimator was proposed by \cite{donoho-johnstone-1994}.
In practice, the shrinkage rule \eqref{rule} is obtained numerically. Figure \ref{fig:rules} shows the shrinkage rule \eqref{rule} under the prior \eqref{prior} for $\sigma = 1$, $\tau = 5$ and for $\alpha \in \{0.6, 0.7, 0.8, 0.9\}$. Note that the rule more severely shrinks large values of $\alpha$. This property is especially important for application at different resolution levels, since finer levels require higher shrinkage of their coefficients.

\vspace{0.3cm}
\centerline{[Figure \ref{fig:rules} around here]}
\vspace{0.3cm}

Thus, we estimate a single wavelet coefficient $\theta$ by applying the shrinkage rule \eqref{rule} on its associated empirical coefficient $z$, i.e
\begin{equation}
    \boldsymbol{\hat{\theta}} = \delta(\boldsymbol{z}). \nonumber
\end{equation}
\noindent Finally, we apply the inverse discrete wavelet transform (IDWT) of the estimated wavelet coefficients vector $\boldsymbol{\hat{\theta}}$ to estimate the function values $f(x_i)$. Since the wavelet transform is orthogonal, the IDWT can be represented by the transpose matrix $\boldsymbol{W^{t}}$ and the estimator of $f$ is calculated by
\begin{equation}
    \boldsymbol{\hat{f}} = \boldsymbol{W^{t}} \boldsymbol{\hat{\theta}}. \nonumber
\end{equation}

\subsection{An example with simulated data}
\begin{figure}
    \centering
    \includegraphics[width=1\linewidth]{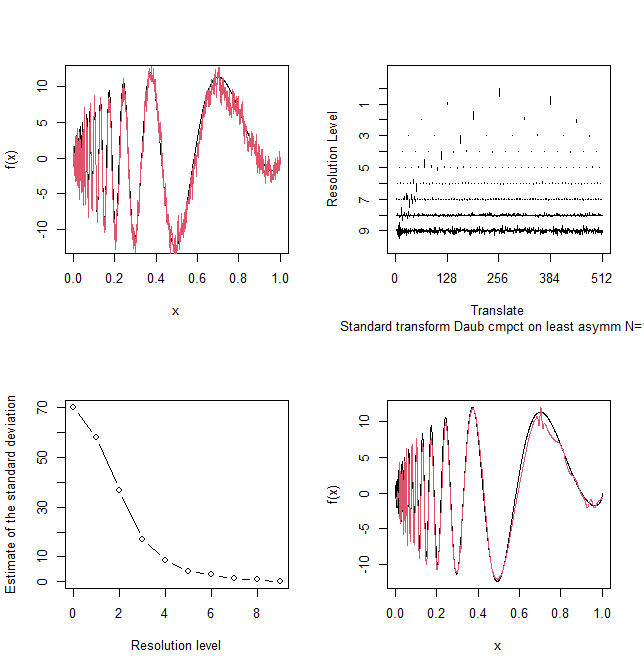}
    \caption{Generated data ($n = 1024$) from the Doppler function with random noise according to an ARFIMA(0,0.4,0) process (top - left). Empirical wavelet coefficients by resolution level (top - right). Estimates of the standard deviation of the empirical coefficients also by resolution level (bottom - left). Estimated curve by the application of the shrinkage rule \eqref{rule} under prior \eqref{prior} (bottom - right).}
    \label{fig:example}
\end{figure}

We illustrate the estimation process by an example involving a Donoho-Johnstone test function called Doppler as underlying function. From model (\ref{timemodel}) we generated $n = 1024 = 2^{10}$ equally spaced points using the Doppler function and added random noise following the ARFIMA(0,0.4,0) process (\ref{eq:model-arfima}). The variance of the process was chosen to obtain a signal-to-noise ratio (SNR) of $5$. Figure \ref{fig:example} (top - left) shows the Doppler function and the generated data. A DWT under a Daubechies basis with ten null moments was applied to the data according to \eqref{waveletmodel} and the empirical wavelet coefficients are represented by resolution level in Figure \ref{fig:example} (top - right). Estimates of the standard deviation of the empirical coefficients by resolution level were calculated and are displayed in Figure \ref{fig:example} (bottom - left). As noted by \cite{johnstone-silerman-1997}, the standard deviation of the empirical coefficients typically decreases when the resolution level increases, which suggests that the parameters of the shrinkage rule should be adjusted by resolution level under stationary processes. Finally we applied the shrinkage rule \eqref{rule} under the prior \eqref{prior} with $\tau = 5$ and $\alpha$ according to \eqref{eq:alpha} on the empirical coefficients and the IDWT to obtain the estimated function, which is shown in Figure \ref{fig:example} (bottom - right). 

\begin{figure}
    \centering
    \includegraphics[width=1\linewidth]{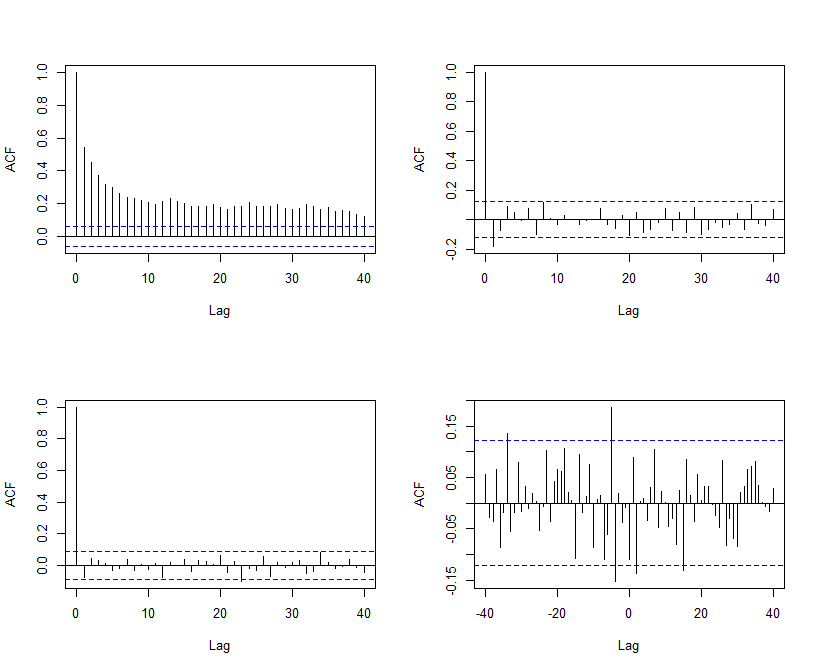}
    \caption{Autocorrelation function of a generated random noise under ARFIMA(0,0.4,0) process (top - left), empirical coefficients in the resolution levels 8 (top - right) and 9 (bottom - left). Cross-correlation function between empirical coefficients in resolution levels 8 and 9 (bottom - right).}
    \label{fig:excorr}
\end{figure}

The method's success in terms of the decorrelation property of the DWT can be viewed in Figure \ref{fig:excorr}. The autocorrelation function of the generated ARFIMA (0,0.4,0) random noise is shown in Figure \ref{fig:excorr} (top - left). In fact, we observe significant autocorrelations until the lag equals 40. On the other hand, Figure \ref{fig:excorr} (top - right) and (bottom - left) show the autocorrelation function of the empirical coefficients at resolution levels 8 and 9 respectively, indicating the reduced correlation in the wavelet domain. Furthermore, the cross-correlation function between the empirical coefficients in those resolution levels are represented in Figure \ref{fig:excorr} (bottom - right). Again we see the decorrelation action of the wavelet transformation. This property justifies the application of the shrinkage rule in the estimation process. 

\vspace{0.3cm}
\centerline{[Figures \ref{fig:example} and \ref{fig:excorr} around here]}
\vspace{0.3cm}

\section{Simulation studies}\label{sec:simulations}

We conducted simulation studies to evaluate the performance of the shrinkage rule under logistic prior in nonparametric models with stationary errors. We generated data from model \eqref{timemodel} with errors under the AR(1) process with $\phi = 0.25$, $0.5$ and $0.9$ in \eqref{eq:model-ar1} and under the ARFIMA(0, $d$, 0) process with $d = 0.2$ and $0.4$ in \eqref{eq:model-arfima}. Furthermore, the variances of the generated errors were chosen according to three signal-to-noise ratios (SNR), $\mathrm{SNR} = 3, 5$ and $7$, and three sample sizes $n = 512, 1024$ and $2048$. The goal was to evaluate how good the shrinkage rule works under different correlation structures of the errors, level of noise present in the data, which is controlled by the signal-to-noise ratios, and different sample sizes. The hyperparameters of the rule were adopted according to \eqref{eq:alpha} and $\tau = 5$. We also considered the generation of IID errors, the standard case, to compare the performance with correlated and non-correlated errors.

The four Donoho - Johnstone (D-J) test functions in the interval $[0,1]$ called Bumps, Blocks, Doppler and Heavisine, were considered as underlying functions to be estimated in model \eqref{timemodel}. These functions, which are the benchmark functions in wavelet studies, are defined as follows:

\begin{itemize}

\item[(a)] BUMPS. 
\begin{equation}
f(x) = \sum_{l=1}^{11} h_l K\left(\frac{x - x_l}{w_l} \right),
\end{equation}
where $K(x) = (1 + |x|)^{-4}$ and
\begin{eqnarray*}
(x_l)_{l=1}^{11} & = &(0.1, 0.13, 0.15, 0.23, 0.25, 0.40, 0.44, 0.65, 0.76, 0.78, 0.81)\\
(h_l)_{l=1}^{11} & = &(4, 5, 3, 4, 5, 4.2, 2.1, 4.3, 3.1, 5.1, 4.2)\\
(w_l)_{l=1}^{11} & = &(0.005, 0.005, 0.006, 0.01, 0.01, 0.03, 0.01, 0.01, 0.005, 0.008, 0.005)
\end{eqnarray*}

\item[(b)] BLOCKS.
\begin{equation}
f(x) = \sum_{l=1}^{11} h_l K(x - x_l),
\end{equation}
where $K(x) = (1 + \mathrm{sgn}(x))/2$ and
\begin{eqnarray*}
(x_l)_{l=1}^{11} & = & (0.1, 0.13, 0.15, 0.23, 0.25, 0.40, 0.44, 0.65, 0.76, 0.78, 0.81)  \\
(h_l)_{l=1}^{11} & = &(4, -5, 3, -4, 5, -4.2, 2.1, 4.3, -3.1, 2.1, -4.2) 
\end{eqnarray*}

\item[(c)] DOPPLER.
\begin{equation}
f(x) = \sqrt{x(1-x)}\sin\left(\frac{2.1 \pi}{x + 0.05} \right).    
\end{equation}

\item[(d)] HEAVISINE.
\begin{equation}
f(x) = 4\sin(4 \pi x) - \mathrm{sgn}(x - 0.3) - \mathrm{sgn}(0.72 - x).
\end{equation}

\end{itemize}

\begin{figure}
\centering
\includegraphics[scale=0.80]{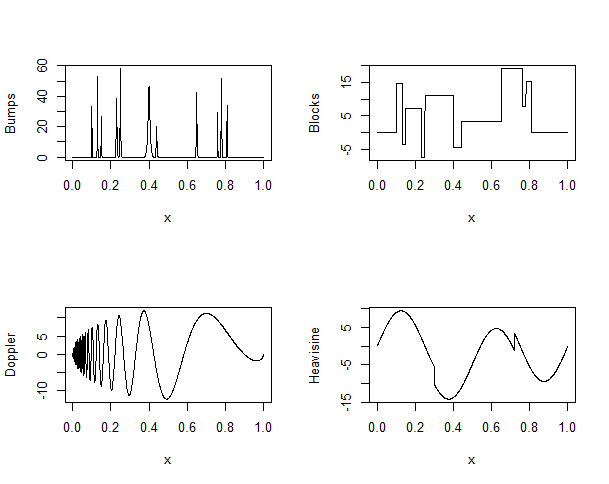}
\caption{ The four Donoho-Johnstone test functions called Bumps, Blocks, Doppler and Heavisine considered as underlying functions in the simulation studies.}\label{fig:functions}
\end{figure}

Plots of the D-J test functions are depicted in Figure \ref{fig:functions}. These functions have important features to be recovered in the wavelet estimation process such as peaks (Bumps), discontinuities (Blocks), oscillations (Doppler) and cusps (Heavisine). 

\vspace{0.3cm}
\centerline{[Figure \ref{fig:functions} around here]}
\vspace{0.3cm}

For each scenario, defined as the combination of underlying function, noise distribution, sample size and signal to noise ratio, $200$ replications were generated. We used the mean squared error (MSE) as the performance measure. Thus, for each generated replication we computed:
$$\mathrm{MSE}^{(m)} = \frac{1}{n} \sum_{i=1}^{n}[{\hat f^{(m)}(x_i)} - f(x_i)]^2, \nonumber$$
where $\hat f^{(m)}(\cdot)$ is the estimate of the function at a particular point in the $m$-th replication, $m = 1, \cdots, 200$. Then, for each scenario the mean (AMSE), standard deviation, median and interquartile range of $\mathrm{MSE}^{(1)}, \ldots, \mathrm{MSE}^{(200)}$ were calculated.

Tables \ref{tab:Bumps}, \ref{tab:Blocks}, \ref{tab:doppler} and \ref{tab:heavisine} contain the results for the four underlying functions Bumps, Blocks, Doppler and Heavisine, respectively. Overall, the performance of the shrinkage rule under logistic prior was similar for all the underlying functions. Thus, for fixed sample size and SNR,  the mean and the standard deviation of the MSEs increased as the parameter $\phi$ increased, under AR errors. The same occured under ARFIMA errors when the parameter $d$ moved from 0.2 to 0.4. Therefore, as expected,  the precision of the estimates decreased as the  correlation became stronger. Moreover, the means of the MSEs were bigger for AR errors with $\phi=0.9$ than for ARFIMA(0,0.4,0) errors and the standard deviations were smaller for the former in all  scenarios\footnote{These findings also held when we considered the median instead the mean and the interquartile range instead the standard deviation, see the Supplementary Material.}. Thus, among the considered error models, the shrinkage rule worked worse, compared to the IID case, for estimating functions under AR errors with $\phi=0.9$. In addition the signal-to-noise ratio had a strong impact on the performance of the shrinkage rule. For example, under the Bumps function in Table \ref{tab:Bumps}, $n = 512$ and AR errors with $\phi=0.9$, the mean of the MSE was $5.062$ for $\mathrm{SNR} = 3$ and $1.144$ for $\mathrm{SNR} = 7$. Similar conclusions were obtained based on the medians, see the Supplementary Material. Besides, the standard deviations became smaller as the SNR increased, for fixed sample size and noise process. 

\vspace{0.3cm}
\centerline{[Tables \ref{tab:Bumps}, \ref{tab:Blocks} \ref{tab:doppler} and \ref{tab:heavisine} around here]}
\vspace{0.3cm}

\begin{figure}
    \centering
    \includegraphics[width=1\linewidth]{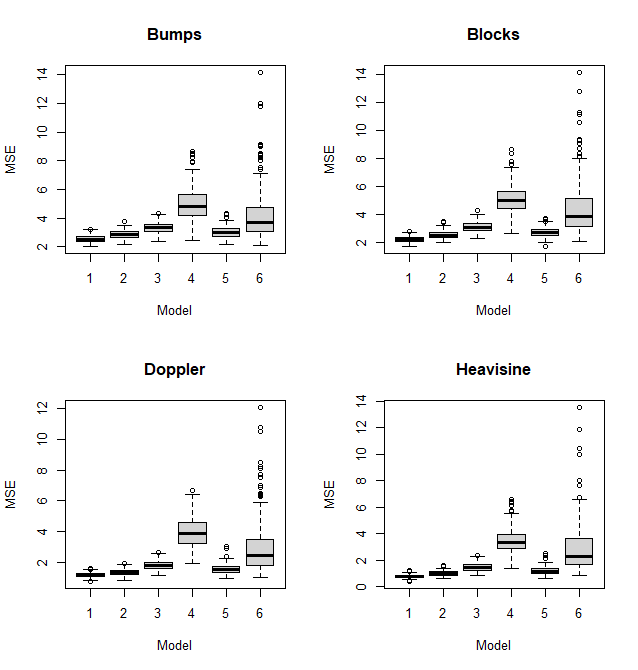}
    \caption{Boxplots of the MSEs of the shrinkage rule under logistic prior according to the errors: 1-iid, AR(1) with 2-$\phi=0.25$, 3-$\phi=0.5$ and 4-$\phi=0.9$ and ARFIMA with 5-$d=0.2$ and 6-$d=0.4$. Results for the Donoho-Johnstone test underlying functions with $n=512$ and $\mathrm{SNR} = 3$.
    }
    \label{fig:bp1}
\end{figure}

Figure \ref{fig:bp1} shows the boxplots of the MSEs of the shrinkage rule under logistic prior for $n = 512$ and $\mathrm{SNR = 3}$. Here we observe that the higher MSEs occurred under AR with $\phi = 0.9$ and ARFIMA(0,0.4,0) error noises. Indeed, the highest variability occurred in those scenarios. In addition, note that the MSEs under AR with $\phi = 0.25$ and $\phi = 0.5$ and ARFIMA(0,0.4,0) errors are closer than those of the IID case compared to the AR with $\phi = 0.9$ and ARFIMA(0,0.4,0) errors. Table \ref{tab:ratio} reports the ratios between the mean of the MSEs under the correlated noises over the mean corresponding to the IID case for the underlying Bumps function. Here, for fixed sample size and SNR, the rates increased with stronger correlation. Thus, in the AR class, the rates increased as $\phi$ increased and the same happened in the ARFIMA class when $d$ increases. In addition, for fixed SNR and error noise process, we observed a small increase in the rates as the sample size increased, except for the ARFIMA(0,0.4,0) case where the increase was moderate. This was because the AMSE of the rule under ARFIMA(0,0.4,0) errors decreased more slowly than its AMSE under IID errors as the sample size increased. To illustrate this, for the Heavisine function and $\mathrm{SNR = 3}$, the AMSE under IID errors decreased from 0.799 ($n = 512$) to 0.602 ($n = 2048$), a reduction of $24.6 \%$, while under ARFIMA(0,0.4,0) errors, the AMSE declined from 3.015 ($n = 512$) to 2.831 ($n = 2048$), a reduction of $6.1 \%$.  Moreover, for fixed sample size and error noise process the ratios were similar across the SNR values. For instance, for $n = 2048$, the ratios under ARFIMA(0,0.4,0) errors were 2.61, 2.58 and 2.69 for $\mathrm{SNR = 3}$, $5$ and $9$ respectively. Qualitatively similar results were obtained for the Blocks, Doppler and Heavisine functions, see the Supplementary Material. 

\vspace{0.3cm}
\centerline{[Figure \ref{fig:bp1} around here]}
\vspace{0.3cm}

\begin{figure}
    \centering
    \includegraphics[width=1\linewidth]{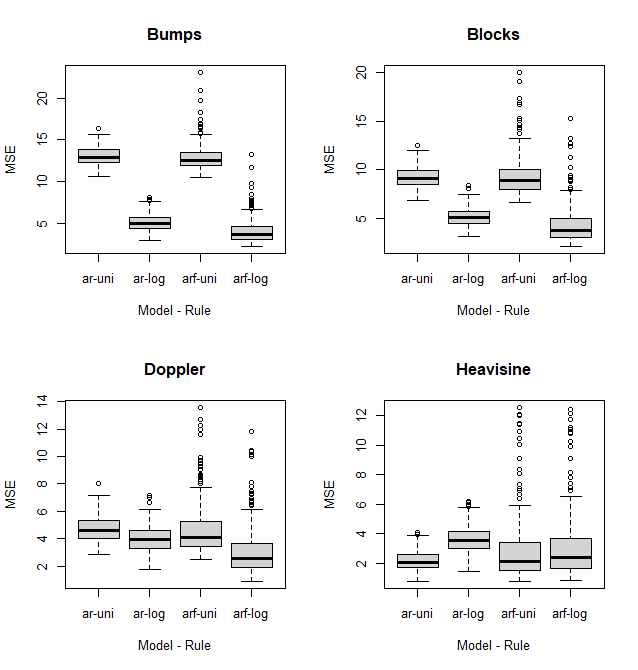}
    \caption{Boxplots of the MSEs of shrinkage rule under logistic prior according to AR(1) with $\phi=0.9$ (ar-log) and ARFIMA(0,0.4,0) (arf-log) errors and the MSEs of the soft thresholding rule with level-dependent universal threshold \eqref{softrule} according to the same respective errors structures (ar-uni and arf-uni). Results of the Donoho-Johnstone test of underlying functions with $n=512$ and $\mathrm{SNR} = 3$.}
    \label{fig:bp2}
\end{figure}

Furthermore, we compared the performance of the proposed method with the standard soft thresholding rule $\eta_S(\cdot)$ proposed by \cite{donoho-johnstone-1994},
\begin{equation}\label{softrule}
\eta_S(z) = \begin{cases} 
 0, & \text{if $|z| \leq \lambda$} \\  
 \mathrm{sgn}(z)(|z| - \lambda), & \text{if $|z|>\lambda$,}   
 \end{cases} 
\end{equation}
where $\lambda > 0$ is a threshold value and $\mathrm{sgn}(z)$ represents the sign of $z$. The threshold value $\lambda$ in \eqref{softrule} was chosen to be the level-dependent universal threshold according to \cite{johnstone-silerman-1997} for correlated noise,
\begin{equation}
\lambda_j = \hat{\sigma}_j \sqrt{2\log(n)}, \nonumber
\end{equation}
where $\hat{\sigma}_j$ is obtained from \eqref{sigma}.  
Figure \ref{fig:bp2} shows the results for $n=512$ and SNR$=3$ with errors following AR with $\phi = 0.9$ and ARFIMA(0,0.4,0) processes. Comparison of the boxplots of the MSE's shows that the logistic shrinkage rule worked clearly better than the thresholding rule \eqref{softrule} for the Bumps and Blocks functions for both error structures considered. Under the Doppler function, the proposed rule was still better than the thresholding rule, even though these methods had almost the same performance for AR errors with $\phi = 0.9$. For the Heavisine function, the rules had similar results for ARFIMA(0,0.4,0) errors but the soft thresholding was better for AR with $\phi = 0.9$ errors. The same conclusions were obtained for the remaining  scenarios (combinations of sample sizes and signal-to-noise ratios). Therefore, in general, the logistic shrinkage rule performed better than the soft thresholding rule under the level-dependent universal threshold. 

\vspace{0.3cm}
\centerline{[Figure \ref{fig:bp2} around here]}
\vspace{0.3cm}

\section{Illustration}\label{sec:illustration}
We applied the shrinkage rule under logistic prior to estimate the light curve for the variable star RU Andromeda. The dataset consisted of $n = 256$ magnitudes collected at irregularly spaced times from Julian Day 2,440,043 to 2,441,592, corresponding to July 5, 1968 to October 1, 1972. The data were gathered from the international database of the American Association of Variable Star Observers (AAVSO)  at \textit{www.aavso.org}. \cite{sardy-et-al-1999} analyzed this dataset with uncorrelated errors and \cite{porto-et-al-2016} took the correlated errors into account and applied the thresholding rule \eqref{softrule} to estimate the light curve. 

\begin{figure}
\centering
\includegraphics[scale=0.70]{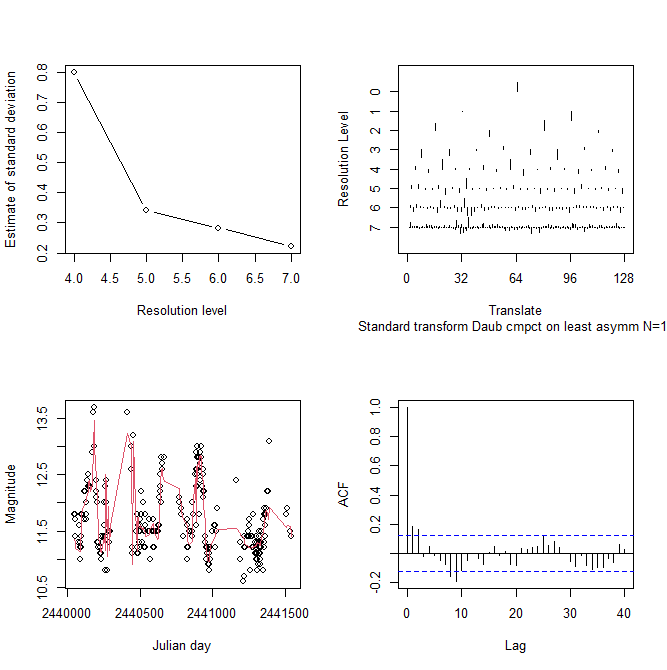}
\caption{Estimates of the standard deviations of the errors by resolution level for the variable star RU Andromeda dataset application. The primary resolution level adopted was $J_0 = 4$ (top-left), empirical wavelet coefficients by resolution level (top-right), dataset ($n=512$) of light magnitudes of the variable star RU Andromeda considered in the application and the estimated light curve (in red) by the shrinkage rule under logistic prior (bottom-left) and autocorrelation function of the estimated errors in time domain (bottom-right).}\label{fig:application}
\end{figure}

Figure \ref{fig:application} (top-left) shows the estimates of the standard deviation of the errors according to \eqref{sigma} from the primary resolution level ($J_0 = 4$) until the highest resolution level ($J = 7$).  The estimates decreased as the resolution level increased, which usually occurs in wavelet domains with correlated noises. The empirical wavelet coefficients after the application of a DWT under the Daubechies basis with ten null moments are depicted in Figure \ref{fig:application} (top-right). The significant coefficients are in the coarser resolution levels. In this sense, most of the coefficients at finer resolution levels should be shrunk more drastically. 

The original dataset contains two or even three measures of magnitudes on some Julian days. In these cases we considered the median of the magnitudes, as in \cite{porto-et-al-2016}. Figure \ref{fig:application} (bottom-left) provides the data considered and the light curve estimate by the shrinkage rule under logistic prior. The estimated curve captures the signal of the data. Moreover, the estimated curve has peak magnitudes on the Julian days 2,440,183 (November 22, 1968); 2,440,415 (July 12, 1969); 2,440,452 (August 18, 1969) and 2,441,394 (March 17, 1972), with magnitudes greater than 13.0. 
Finally, Figure \ref{fig:application} (bottom-right) shows the autocorrelation function of the estimated errors. Although there is almost no correlation structure, we rejected the hypothesis of non-correlation among the residuals according to the Box-Ljung (BL) test based on 10 autocorrelations. The observed BL test statistic is 38.846 with associated p-value less than 0.001. This indicates the presence of a nonzero correlation structure in the errors. 

\vspace{0.3cm}
\centerline{[Figure \ref{fig:application} around here]}
\vspace{0.3cm}

\section{Conclusions}\label{sec:conclusions}
In the present work, we proposed a Bayesian rule based on the mixture of a point mass function at zero and the logistic distribution to perform wavelet shrinkage in nonparametric regression models with stationary errors. The rule is an adaptation of \cite{sousa-2020}, which is level-dependent, i.e, the severity of the shrinkage depends on the resolution level of the coefficients. The hyperparameters of the prior distribution also control the shrinkage level of the rule, which facilitated their elicitation. 

Simulation studies indicated that the precision of the estimates decreased as the correlation level increased. In addition, given a sample size and error correlated noise, the performance of the rule was almost the same as the SNR decreased, compared to the performance of the rule under IID errors. Furthermore, under AR(1) with $\phi = 0.9$ and ARFIMA(0,0.4,0) errors, the performance of the proposal was better than that of the standard soft thesholding rule with universal policy in most of the considered underlying functions, sample sizes and signal-to-noise ratios scenarios. This suggests that the proposed procedure is a useful alternative to perform wavelet shrinkage under correlated errors with short or long memory processes. 

The impact of the chosen wavelet basis on the performance of the shrinkage rule and the proposition of Bayesian rules that act asymmetrically in the shrinkage process under correlated errors are suggestions for possible future works.


\bibliographystyle{plainnat}
\bibliography{manuscript}

\begin{thebibliography}{19}
\providecommand{\natexlab}[1]{#1}
\providecommand{\url}[1]{\texttt{#1}}
\expandafter\ifx\csname urlstyle\endcsname\relax
  \providecommand{\doi}[1]{doi: #1}\else
  \providecommand{\doi}{doi: \begingroup \urlstyle{rm}\Url}\fi

\bibitem[Angelini and Vidakovic(2004)]{angelini-vidakovic-2004}
C.~Angelini and B.~Vidakovic.
\newblock Gama-minimax wavelet shrinkage: a robust incorporation of information about energy of a signal in denoising applications.
\newblock \emph{Stat Sin}, 14\penalty0 (1):\penalty0 103--125, 2004.

\bibitem[Chipman et~al.(1997)Chipman, Kolaczyk, and McCulloch]{Chipman-et-al-1997}
H.~Chipman, E.~Kolaczyk, and R.~McCulloch.
\newblock Adaptive bayesian wavelet shrinkage.
\newblock \emph{J Am Stat Assoc}, 92\penalty0 (1):\penalty0 1413--1421, 1997.

\bibitem[Daubechies(1992)]{daubechies-1992}
I.~Daubechies.
\newblock \emph{Ten lectures on wavelets}.
\newblock SIAM, Philadelphia (GA), 1992.

\bibitem[Donoho and Johnstone(1994)]{donoho-johnstone-1994}
D.L. Donoho and I.M. Johnstone.
\newblock Ideal spatial adaptation by wavelet shrinkage.
\newblock \emph{Biometrika}, 81\penalty0 (1):\penalty0 425--455, 1994.

\bibitem[Donoho and Johnstone(1995)]{donoho-johnstone-1995}
D.L. Donoho and I.M. Johnstone.
\newblock Adapting to unknown smoothness via wavelet shrinkage.
\newblock \emph{J Am Stat Assoc}, 90\penalty0 (1):\penalty0 1200--1224, 1995.

\bibitem[Granger and Joyeux(1980)]{granger-joyeux-1980}
C.~W.~J. Granger and R.~Joyeux.
\newblock An introduction to long‐memory time series models and fractional differencing.
\newblock \emph{Journal of Time Series Analysis}, 1\penalty0 (1):\penalty0 15--29, 1980.

\bibitem[Hosking(1981)]{hosking-1981}
J.R.M. Hosking.
\newblock Fractional differencing.
\newblock \emph{Biometrika}, 68\penalty0 (1):\penalty0 165--176, 1981.

\bibitem[Johnstone and Silverman(1997)]{johnstone-silerman-1997}
I.M. Johnstone and B.W. Silverman.
\newblock Wavelet threshold estimators for data with correlated noise.
\newblock \emph{J. R. Statist. Soc. B}, 59\penalty0 (2):\penalty0 319--351, 1997.

\bibitem[Mallat(1998)]{mallat-1998}
S.G. Mallat.
\newblock \emph{A wavelet tour of signal processing}.
\newblock Academic Press, San Diego (CA), 1998.

\bibitem[Porto et~al.(2016)Porto, Morettin, Percival, and Aubin]{porto-et-al-2016}
R.~Porto, P.A. Morettin, D.~Percival, and E.~Aubin.
\newblock Wavelet shrinkage for regression models with random design and correlated errors.
\newblock \emph{Brazilian Journal of Probability and Statistics}, 30\penalty0 (4):\penalty0 614--652, 2016.

\bibitem[Reményi and Vidakovic(2015)]{remenyi-vidakovic-2015}
N.~Reményi and B.~Vidakovic.
\newblock Wavelet shrinkage with double weibull prior.
\newblock \emph{Commun Stat Simul Comput}, 44\penalty0 (1):\penalty0 88--104, 2015.

\bibitem[Robert(2007)]{robert-2007}
P.R. Robert.
\newblock \emph{The Bayesian Choice: From Decision-Theoretic Foundations to Computational Implementation}.
\newblock Springer, New York, 2007.

\bibitem[Sardy et~al.(1999)Sardy, Percival, Bruce, Gao, and Stuetzle]{sardy-et-al-1999}
S.~Sardy, D.~Percival, A.G. Bruce, H.~Gao, and W.~Stuetzle.
\newblock Wavelet shrinkage for unequally spaced data.
\newblock \emph{Stat. Comput}, 9\penalty0 (1):\penalty0 65--75, 1999.

\bibitem[Sousa(2020)]{sousa-2020}
A.R.S. Sousa.
\newblock Bayesian wavelet shrinkage with logistic prior.
\newblock \emph{Commun Stat Simul Comput}, 51\penalty0 (8):\penalty0 4700--4714, 2020.

\bibitem[Sousa et~al.(2020)Sousa, Garcia, and Vidakovic]{sousa-et-al-2020}
A.R.S. Sousa, N.L. Garcia, and B.~Vidakovic.
\newblock Bayesian wavelet shrinkage with beta prior.
\newblock \emph{Computational Statistics}, 36\penalty0 (2):\penalty0 1341--1363, 2020.

\bibitem[Vidakovic(1999)]{vidakovic-1999}
B.~Vidakovic.
\newblock \emph{Statistical modeling by wavelets}.
\newblock Wiley, New York, 1999.

\bibitem[Vidakovic and Ruggeri(2001)]{vidakovic-ruggeri-2001}
B.~Vidakovic and F.~Ruggeri.
\newblock Bams method: theory and simulations.
\newblock \emph{Sankhya Indian J Stat B}, 63\penalty0 (1):\penalty0 234--249, 2001.

\bibitem[Vimalajeewa et~al.(2023)Vimalajeewa, Dasgupta, Ruggeri, and Vidakovic]{vimalajeewa-et-al-2023}
D.~Vimalajeewa, A.~Dasgupta, F.~Ruggeri, and B.~Vidakovic.
\newblock Gamma-minimax wavelet shrinkage for signals with low snr.
\newblock \emph{The New England Journal of Statistics in Data Science}, 0\penalty0 (1):\penalty0 1--13, 2023.

\bibitem[Wang(1996)]{wang-1996}
Y.~Wang.
\newblock Function estimation via wavelet shrinkage for long-memory data.
\newblock \emph{The Annals of Statistics}, 24\penalty0 (2):\penalty0 466--484, 1996.

\end{thebibliography}


\clearpage
\begin{table}
\small
    \centering
    \begin{tabular}{c l c c c}\hline
        $\boldsymbol{n}$ & \textbf{Noise process} & $\boldsymbol{\mathrm{SNR} = 3}$  & $\boldsymbol{\mathrm{SNR} = 5}$  & $\boldsymbol{\mathrm{SNR} = 7}$ \\ \hline \hline
       512  & IID  & 2.580 (0.226)  & 1.040 (0.092)   & 0.599 (0.050) \\
            & AR(1) ($\phi = 0.25$)  & 2.833 (0.298)  & 1.176 (0.109)  & 0.670 (0.055)\\
            & AR(1) ($\phi = 0.5$) & 3.369 (0.419) & 1.381 (0.138) & 0.791 (0.079) \\
            & AR(1) ($\phi = 0.9$) & 5.062 (1.048) &  2.041 (0.382) & 1.144 (0.211) \\
            & ARFIMA(0,0.2,0) &  2.970 (0.354) &  1.243 (0.140) & 0.711 (0.071) \\
            & ARFIMA(0,0.4,0) & 4.395 (2.032)    & 1.802 (0.792) & 0.973 (0.327)\\ 
    \\[-0.5em]

            1024 & IID & 1.832 (0.136) & 0.731 (0.048)& 0.406 (0.029) \\
                 & AR(1) ($\phi = 0.25$) & 2.121 (0.184) & 0.861 (0.072) & 0.485 (0.043) \\
                 & AR(1) ($\phi = 0.5$) & 2.703 (0.246) & 1.073 (0.094) & 0.601 (0.055) \\
                 & AR(1) ($\phi = 0.9$) & 4.613 (0.659) & 1.796 (0.248) & 0.975 (0.141) \\
                 & ARFIMA(0,0.2,0) & 2.257 (0.234) & 0.931 (0.104) & 0.513 (0.045) \\
                 &ARFIMA(0,0.4,0) & 3.749 (1.866) & 1.545 (0.594) & 0.807 (0.317) \\
    \\[-0.5em]
        2048 & IID & 1.261 (0.080) & 0.475 (0.031) & 0.264 (0.017)\\
             & AR(1) ($\phi = 0.25$) & 1.527 (0.106) & 0.576 (0.043)& 0.321 (0.021)\\
             & AR(1) ($\phi = 0.5$) & 2.084 (0.182)  & 0.790 (0.056)& 0.434 (0.036) \\
             & AR(1) ($\phi = 0.9$) & 4.029 (0.510)   &1.578 (0.187) & 0.843 (0.092) \\
             & ARFIMA(0,0.2,0) & 1.740 (0.161)& 0.665 (0.665)  & 0.363 (0.032) \\
             & ARFIMA(0,0.4,0) & 3.285 (1.621) & 1.227 (0.539)& 0.709 (0.301)\\
    \hline             
    \end{tabular}
    \caption{Mean (standard deviation) of the MSEs of the shrinkage rule under logistic prior with the underlying Bumps function in the simulation studies.}
    \label{tab:Bumps}
\end{table}

\begin{table}
\small
    \centering
    \begin{tabular}{c l c c c}\hline
        $\boldsymbol{n}$ & \textbf{Noise process} & $\boldsymbol{\mathrm{SNR} = 3}$  & $\boldsymbol{\mathrm{SNR} = 5}$  & $\boldsymbol{\mathrm{SNR} = 7}$ \\ \hline \hline 
       512 & IID & 2.246 (0.199)  & 0.937 (0.098) & 0.506 (0.053) \\
           & AR(1) ($\phi = 0.25$) & 2.532 (0.265) & 1.072 (0.112)& 0.589 (0.058) \\
           & AR(1) ($\phi = 0.5$) & 3.158 (0.383) &1.318 (0.137)& 0.712 (0.076) \\
           & AR(1) ($\phi = 0.9$) & 5.329 (1.052) & 2.033 (0.359)&1.078 (0.189) \\
           & ARFIMA(0,0.2,0) & 2.772 (0.327) & 1.166 (0.145)&0.626 (0.077) \\
           & ARFIMA(0,0.4,0) & 4.617 (2.274) & 1.724 (0.795)&0.923 (0.366) \\
        \\[-0.5em]
    1024 & IID & 1.537 (0.124) & 0.613 (0.052) & 0.354 (0.030) \\
         & AR(1) ($\phi = 0.25$) & 1.795 (0.150) & 0.727 (0.066) & 0.415 (0.036) \\
         & AR(1) ($\phi = 0.5$) & 2.407 (0.263) & 0.959 (0.103) & 0.530 (0.048) \\
         & AR(1) ($\phi = 0.9$) & 4.421 (0.760) & 1.731 (0.257) & 0.938 (0.131) \\
         & ARFIMA(0,0.2,0) & 2.020 (0.221) & 0.823 (0.090) & 0.465 (0.046) \\
         & ARFIMA(0,0.4,0) & 3.510 (1.493) & 1.453 (0.673) & 0.766 (0.304) \\
    \\[-0.5em]
    2048 & IID & 1.163 (0.084) & 0.439 (0.032) & 0.241 (0.016) \\
         & AR(1) ($\phi = 0.25$) & 1.390 (0.109) & 0.535 (0.037) & 0.292 (0.022) \\
         & AR(1) ($\phi = 0.5$) & 1.921 (0.170) & 0.746 (0.065) & 0.406 (0.032) \\
         & AR(1) ($\phi = 0.9$) & 3.922 (0.522) & 1.522 (0.184) & 0.817 (0.088) \\
         & ARFIMA(0,0.2,0) & 1.601 (0.150) & 0.632 (0.066) & 0.342 (0.032) \\
         & ARFIMA(0,0.4,0) & 3.181 (1.370) & 1.306 (0.699) & 0.685 (0.303) \\
         \hline           
    \end{tabular}
    \caption{Mean (standard deviation) of the MSEs of the shrinkage rule under logistic prior with the underlying Blocks function in the simulation studies.}
    \label{tab:Blocks}
\end{table}

\begin{table}
\small
    \centering
    \begin{tabular}{c l c c c}\hline
        $\boldsymbol{n}$ & \textbf{Noise process} & $\boldsymbol{\mathrm{SNR} = 3}$  & $\boldsymbol{\mathrm{SNR} = 5}$  & $\boldsymbol{\mathrm{SNR} = 7}$ \\ \hline \hline
    512 & IID & 1.181 (0.154) & 0.465 (0.067) & 0.247 (0.036) \\
        &AR(1) ($\phi = 0.25$) & 1.392 (0.219) & 0.568 (0.080) & 0.297 (0.043) \\
        & AR(1) ($\phi = 0.5$) & 1.874 (0.355) & 0.742 (0.114) & 0.400 (0.067) \\
        & AR(1) ($\phi = 0.9$) & 3.915 (1.012) & 1.496 (0.345) & 0.757 (0.180) \\
        & ARFIMA(0,0.2,0) & 1.611 (0.298) & 0.645 (0.116) & 0.340 (0.059) \\
        & ARFIMA(0,0.4,0) & 3.067 (2.139) & 1.264 (0.774) & 0.700 (0.505) \\
    \\[-0.5em]

    1024 & IID & 0.880 (0.113) & 0.302 (0.038) & 0.164 (0.018) \\
         & AR(1) ($\phi = 0.25$) & 1.073 (0.142) & 0.379 (0.051) & 0.206 (0.026) \\
         & AR(1) ($\phi = 0.5$)  & 1.547 (0.227) & 0.578 (0.094) & 0.302 (0.041) \\
         & AR(1) ($\phi = 0.9$) & 3.540 (0.684) & 1.356 (0.250) & 0.712 (0.119) \\
         & ARFIMA(0,0.2,0) & 1.258 (0.187) & 0.452 (0.074) & 0.254 (0.039) \\
         & ARFIMA(0,0.4,0) & 2.886 (1.985) & 1.062 (0.598) & 0.596 (0.343) \\
    \\[-0.5em]

    2048 & IID & 0.714 (0.069) & 0.218 (0.023) & 0.106 (0.013) \\
         & AR(1) ($\phi = 0.25$) & 0.895 (0.089) & 0.286 (0.030) & 0.140 (0.016) \\
         & AR(1) ($\phi = 0.5$) & 1.369 (0.141) & 0.473 (0.054) & 0.234 (0.029) \\
         & AR(1) ($\phi = 0.9$)& 3.341 (0.501) & 1.255 (0.180) & 0.656 (0.092) \\
         & ARFIMA(0,0.2,0) & 1.050 (0.143) & 0.369 (0.054) & 0.191 (0.029) \\
         & ARFIMA(0,0.4,0) & 2.786 (1.859) & 0.979 (0.505) & 0.515 (0.245) \\
         \hline

    \end{tabular}
    \caption{Mean (standard deviation) of the MSEs of the shrinkage rule under logistic prior with the undrlying Doppler function in the simulation studies.}
    \label{tab:doppler}
\end{table}

\begin{table}
\small
    \centering
    \begin{tabular}{c l c c c }\hline
        $\boldsymbol{n}$ & \textbf{Noise process} & $\boldsymbol{\mathrm{SNR} = 3}$  & $\boldsymbol{\mathrm{SNR} = 5}$  & $\boldsymbol{\mathrm{SNR} = 7}$ \\ \hline \hline
    512 & IID & 0.799 (0.143) & 0.290 (0.050) & 0.164 (0.024) \\
        & AR(1) ($\phi = 0.25$) & 0.979 (0.185) & 0.358 (0.064) & 0.208 (0.037) \\
        & AR(1) ($\phi = 0.5$) & 1.443 (0.288) & 0.550 (0.114) & 0.291 (0.053) \\
        & AR(1) ($\phi = 0.9$) & 3.450 (0.929) & 1.357 (0.379) & 0.714 (0.179) \\
        & ARFIMA(0,0.2,0) & 1.191 (0.302) & 0.446 (0.109) & 0.250 (0.055) \\
        & ARFIMA(0,0.4,0) & 3.015 (2.354) & 1.084 (0.754) & 0.579 (0.374) \\
        \\[-0.5em]

    1024 & IID & 0.686 (0.086) & 0.213 (0.030) & 0.109 (0.017) \\
         & AR(1) ($\phi = 0.25$) & 0.849 (0.133) & 0.280 (0.044) & 0.145 (0.023) \\
         & AR(1) ($\phi = 0.5$)  & 1.315 (0.206) & 0.459 (0.073) & 0.239 (0.040) \\
         & AR(1) ($\phi = 0.9$)  & 3.319 (0.702) & 1.252 (0.263) & 0.638 (0.119) \\
         & ARFIMA(0,0.2,0) & 1.043 (0.231) & 0.362 (0.076) & 0.193 (0.042) \\
         & ARFIMA(0,0.4,0) & 2.712 (1.614) & 0.952 (0.591) & 0.558 (0.320) \\
         \\[-0.5em]

    2048 & IID & 0.602 (0.062) & 0.170 (0.022) & 0.079 (0.010) \\
         & AR(1) ($\phi = 0.25$) & 0.756 (0.087) & 0.229 (0.029) & 0.107 (0.015) \\
         & AR(1) ($\phi = 0.5$) & 1.222 (0.138) & 0.402 (0.053) & 0.199 (0.025) \\
         & AR(1) ($\phi = 0.9$) & 3.188 (0.491) & 1.210 (0.192) & 0.601 (0.093) \\
         & ARFIMA(0,0.2,0) & 0.942 (0.130) & 0.305 (0.055) & 0.156 (0.028) \\
         & ARFIMA(0,0.4,0) & 2.831 (1.739) & 0.928 (0.516) & 0.487 (0.301) \\
    \hline
    \end{tabular}
    \caption{Mean (standard deviation) of the MSEs of the shrinkage rule under logistic prior with the underlying Heavisine  function in the simulation studies.}
    \label{tab:heavisine}
\end{table}

\begin{table}[H]
\centering
\begin{tabular}{c l c c c}
\hline
$\boldsymbol{n}$ & \textbf{Noise process} & $\boldsymbol{\mathrm{SNR} = 3}$  & $\boldsymbol{\mathrm{SNR} = 5}$  & $\boldsymbol{\mathrm{SNR} = 7}$ \\ \hline \hline

512 & IID &    1.00 & 1.00 & 1.00 \\
    & AR(1) ($\phi = 0.25$) & 1.10 & 1.13 & 1.12 \\
    & AR(1) ($\phi = 0.5$) &1.31 & 1.33 & 1.32 \\
    & AR(1) ($\phi = 0.9$) & 1.96 & 1.96 & 1.91 \\
    & ARFIMA(0,0.2,0) & 1.15 & 1.20 & 1.19 \\
    & ARFIMA(0,0.4,0) & 1.70 & 1.73 & 1.62 \\ 
 \\[-0.5em]

1024 & IID & 1.00 & 1.00 & 1.00 \\
     & AR(1) ($\phi = 0.25$) & 1.16 & 1.18 & 1.19 \\
     & AR(1) ($\phi = 0.5$) &  1.48 & 1.47 & 1.48 \\
     & AR(1) ($\phi = 0.9$) & 2.52 & 2.46 & 2.40 \\
     & ARFIMA(0,0.2,0) & 1.23 & 1.27 & 1.26 \\
     & ARFIMA(0,0.4,0) & 2.05 & 2.11 & 1.99 \\
\\[-0.5em]

2048 & IID & 1.00 & 1.00 & 1.00 \\
     & AR(1) ($\phi = 0.25$) & 1.21 & 1.21 & 1.22 \\
     & AR(1) ($\phi = 0.5$)& 1.65 & 1.66 & 1.64 \\
     & AR(1) ($\phi = 0.9$) & 3.20 & 3.32 & 3.19 \\
     & ARFIMA(0,0.2,0) & 1.38 & 1.40 & 1.38 \\
     & ARFIMA(0,0.4,0) & 2.61 & 2.58 & 2.69 \\ \hline
\end{tabular}
\caption{Ratios of the AMSEs of the shrinkage rule under logistic prior for autorregressive and ARFIMA noise processes in relation to the IID noise process for the underlying Bumps function. }
    \label{tab:ratio}
\end{table}


\clearpage

\end{document}